\documentclass[%
reprint,
 amsmath,amssymb,
 aps,
]{revtex4-2}
\usepackage{color}
\usepackage{graphicx}
\usepackage{dcolumn}
\usepackage{bm}
\usepackage{amsmath,amsbsy,amssymb}
\usepackage{mathrsfs}
\usepackage{ulem}
\usepackage{mathcomp}
\usepackage{physics}
\usepackage{comment}
\usepackage{ascmac}
\usepackage{here}

\begin{document}

\preprint{APS/123-QED}

\title{Spin Relaxation and Electrical Control of Spin Transport in Few-Layer $T_{\rm d}$-MoTe$_2$}

\author{T. Wakamura,$^1$ M. Hashisaka,$^2$ M. Bard$^1$, S. Okazaki,$^3$ T. Sasagawa,$^{3, 4}$ T. Taniguchi,$^5$ K. Watanabe,$^6$ K. Muraki,$^1$}
\author{N. Kumada$^1$}
\affiliation{$^1$Basic Research Laboratories, NTT, Inc., 3--1 Morinosato--Wakamiya, Atsugi, 243--0198, Japan}
\affiliation{$^2$Institute for Solid State Physics, University of Tokyo, 5--1--5 Kashiwa--no--ha, Kashiwa 277--8581, Japan}



\affiliation{$^3$Materials and Structures Laboratory, Institute of Science Tokyo, 4259 Nagatsuta, Yokohama, 226--8501, Japan}
\affiliation{$^4$Research Center for Autonomous Systems Materialogy, Institute of Science Tokyo, 4259 Nagatsuta, Yokohama, 226--8501, Japan}

\affiliation{$^5$Research Center for Materials Nanoarchitectonics, National Institute for Materials and Science, 1--1 Namiki, Tsukuba, 305--0044, Japan}

\affiliation{$^6$Research Center for Electronic and Optical Materials, National Institute for Materials and Science, 1--1 Namiki, Tsukuba, 305--0044, Japan}%


\date{\today}

\begin{abstract}
We report electrical tuning of spin relaxation in atomically thin $T_{\rm d}$-MoTe$_2$, a representative low-symmetry transition metal dichalcogenide. Weak antilocalization measurements reveal strong spin--orbit interaction with spin--orbit lengths of tens of nanometers, consistent with recent theoretical predictions. Despite the semimetallic nature of $T_{\rm d}$-MoTe$_2$, in the ultrathin limit the spin--orbit scattering time can be tuned electrostatically by up to threefold. We further investigate vertical electric field effects on the spin--orbit length and time, and demonstrate the modulation of the Rashba parameters. Relation between spin--orbit scattering time and the diffusion coefficient indicates that spin relaxation is governed by the D'yakonov--Perel' mechanism in both monolayer and bilayer $T_{\rm d}$-MoTe$_2$. These results establish few-layer $T_{\rm d}$-MoTe$_2$ as a platform for electrically tunable spin transport, thereby providing insight into spin relaxation in low-symmetry transition metal dichalcogenides.
\end{abstract}

\maketitle



\section{Introduction}

Low-symmetry transition metal dichalcogenides (LS-TMDs) have recently attracted considerable attention in spintronics. Unlike conventional heavy metals such as Pt or W, LS-TMDs exhibit a canted spin Hall effect (SHE), where the spin current, charge current, and spin polarization are $\mathit{not}$ mutually orthogonal \cite{Zhou_2019}. This effect induces an out-of-plane antidamping torque on ferromagnets with perpendicular magnetic anisotropy, offering potential for high-density spintronic applications \cite{Liu_2023}. 

$T_{\rm d}$-MoTe$_2$ is a representative LS-TMD \textcolor{black}{and widely known for its topologically nontrivial band structures. Bulk $T_{\rm d}$-MoTe$_2$ has been theoretically \cite{Soluyanov_2015, Yan_2015} and experimentally \cite{Deng_2016, Jiang_2017} established as a type-II Weyl semimetal, while the quantum spin Hall state has been theoretically proposed for monolayer $T_{\rm d}$-MoTe$_2$ \cite{Qian_2014}. In addition to these intriguing topological properties, $T_{\rm d}$-MoTe$_2$ is also a promising material for spintronics applications.} In bulk $T_{\rm d}$-MoTe$_2$, the canted SHE \cite{Hoque_2021, Zhou_2019, Safeer_2019, Ontoso_2023} and also out-of-plane antidamping torque \cite{MacNeill_2016, MacNeill_2017, Liang_2020, Stiehl_2019} have been experimentally demonstrated. In the few-layer thin films, breaking the glide-plane symmetry along the $c$-axis further induces intriguing spin transport phenomena, such as the planar SHE \cite{Song_2020}.

To fully exploit these functionalities, electrical control of spin transport is essential. In closely related systems with the same crystal structure, such as monolayer (1 L) WTe$_2$, vertical electric fields can induce atomic displacements and the resulting lateral dipole moments, which affect the out-of-plane spin component and activate the D'yakonov--Perel' (DP) spin relaxation channel \cite{Shi_2019, Xu_2024}. These results suggest that similar electric-field control may also be possible in $T_{\rm d}$-MoTe$_2$. Theoretically, for example, spin relaxation in 1 L $T_{\rm d}$-MoTe$_2$ is predicted to be tunable by carrier doping \cite{Vila_2021}. However, such an electrical control of spin transport in $T_{\rm d}$-MoTe$_2$ remains largely unexplored in experiments.
In addition to the technical challenges, there remain fundamental open questions in spin transport in this material, notably a substantial discrepancy between theoretically predicted and experimentally reported spin relaxation lengths, whose origin remains unsolved. For instance, nonlocal measurements in few-layer $T_{\rm d}$-MoTe$_2$ have reported micrometer-scale spin relaxation lengths at room temperature \cite{Song_2020}, which are two orders of magnitude longer than both theoretical predictions for 1 L $T_{\rm d}$-MoTe$_2$ \cite{Vila_2021} and experimental values reported for WTe$_2$ \cite{Zhang_2017}. Such inconsistencies hinder a quantitative understanding of spin transport in $T_{\rm d}$-MoTe$_2$ and call for more robust experimental approaches to evaluate spin--orbit interactions and spin relaxation. 

Weak antilocalization (WAL) measurements provide a reliable and well-controlled method to evaluate spin--orbit interactions and spin relaxation \cite{Niimi_2013}. Whereas WAL signals have been observed in thin $T_{\rm d}$-MoTe$_2$ \cite{Naylor_2016, Wang_2018, Song_2018_WAL}, previous analyses of WAL in thin $T_{\rm d}$-MoTe$_2$ did not explicitly incorporate the spin--orbit length ($L_{\rm so}$) in the fitting procedure. As a result, the evaluation of spin relaxation is indirect, which may partly account for the discrepancies between experimental results and theoretical predictions. Furthermore, the dominant spin relaxation mechanism has not been clearly established in previous experimental studies. 



Here, we present a systematic study of spin relaxation and electrical control of spin transport in few-layer $T_{\rm d}$-MoTe$_2$ (1--6 layers), using low-temperature magnetotransport measurements. By explicitly including $L_{\rm so}$ for fitting WAL signals, we directly evaluate the spin relaxation. The extracted $L_{\rm so}$ is on the order of several tens of nanometers, consistent with theoretical estimates \cite{Vila_2021}. In the ultrathin limit, we demonstrate electrostatic gate control of spin transport, achieving a threefold modulation of the spin relaxation time ($\tau_{\rm so}$). From the relation between $\tau_{\rm so}$ and the diffusion coefficient, we identify the DP mechanism as the dominant spin relaxation process in both 1 L and 2 L samples. Furthermore, dual-gate measurements allow us to isolate the effects of vertical electric fields, revealing variations in $\tau_{\rm so}$ and the Rashba parameters. 

The present paper is organized as follows. Section II describes the experimental methods. Section III presents the experimental results and analyses, including weak antilocalization for different thicknesses (Sec. III A), gate control of spin transport (Sec. III B), identification of spin relaxation mechanisms (Sec. III C) and vertical electric field effects on spin relaxation (Sec. III D). Section IV summarizes the main results of this work.



\begin{figure}[tb!]
\begin{center}
\includegraphics[width=7.5cm,clip]{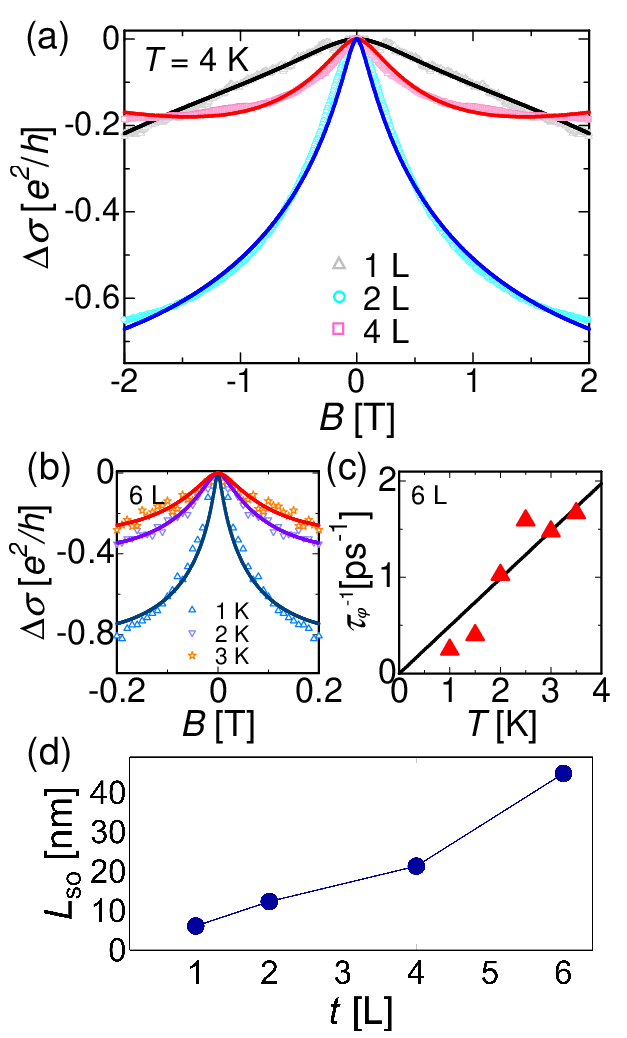}
\caption{(a) Magnetoconductivity variation [$\Delta \sigma (B)$] at 4 K obtained from 1 L, 2 L and 4 L $T_{\rm d}$-MoTe$_2$. Solid curves are fits by using the ILP formula. (b) $\Delta \sigma(B)$ for a 6 L sample at different temperatures with the fits using the ILP formula. (c) $1/\tau_\varphi$ obtained from the fits in (b) as a function of temperature ($T$). (d) Thickness ($t$) dependence of the spin--orbit length ($L_{\rm so}$). The data are obtained from the fits at 4 K (1 L -- 4 L) and 3.5 K (6 L). }
\label{fig1}
\end{center}
\end{figure} 

\section{Methods}


Few-layer $T_{\rm d}$-MoTe$_2$ flakes are mechanically exfoliated \textcolor{black}{on a Si/SiO$_2$ chip} inside a glovebox (O$_2$ and H$_2$O $<$ 0.5 ppm) from high-quality crystals grown by the flux method, with a residual-resistivity ratio (RRR) of $\sim 1000$. The exfoliated flakes are assembled into devices using a standard dry-transfer technique with \textcolor{black}{a polycarbonate (PC) film on a polydimethylsiloxane (PDMS) polymer} and hBN \cite{Zomer_2014, Pizzocchero_2016}, and are placed on prefabricated Au/Ti or Pt/Ti electrodes on SiO$_2$ or hBN substrates. The top hBN layer serves as a protective encapsulation, preventing degradation of $T_{\rm d}$-MoTe$_2$ in ambient conditions \cite{Gan_2020, Yang_2018}. For dual-gated devices, few-layer graphene is used as gate electrodes with hBN as the dielectric. Immediately after fabrication, the samples are mounted in a $^3$He cryostat, sealed in a vacuum can, and cooled down. Electrical transport measurements are performed using a lock-in technique at a frequency of 77 Hz. The flake thickness is determined by atomic force microscopy after the transport measurements. In this study, we analyze data from one 1 L, three 2 L, three 4 L, and one 6 L samples. Further details of the fabrication process are provided in Refs.~\cite{Wakamura_2024, Wakamura_2026}.
\section{Measurement and analysis}
\subsection{Weak antilocalization for different thicknesses}

\begin{figure*}[tb!]
\begin{center}
\includegraphics[width=15cm,clip]{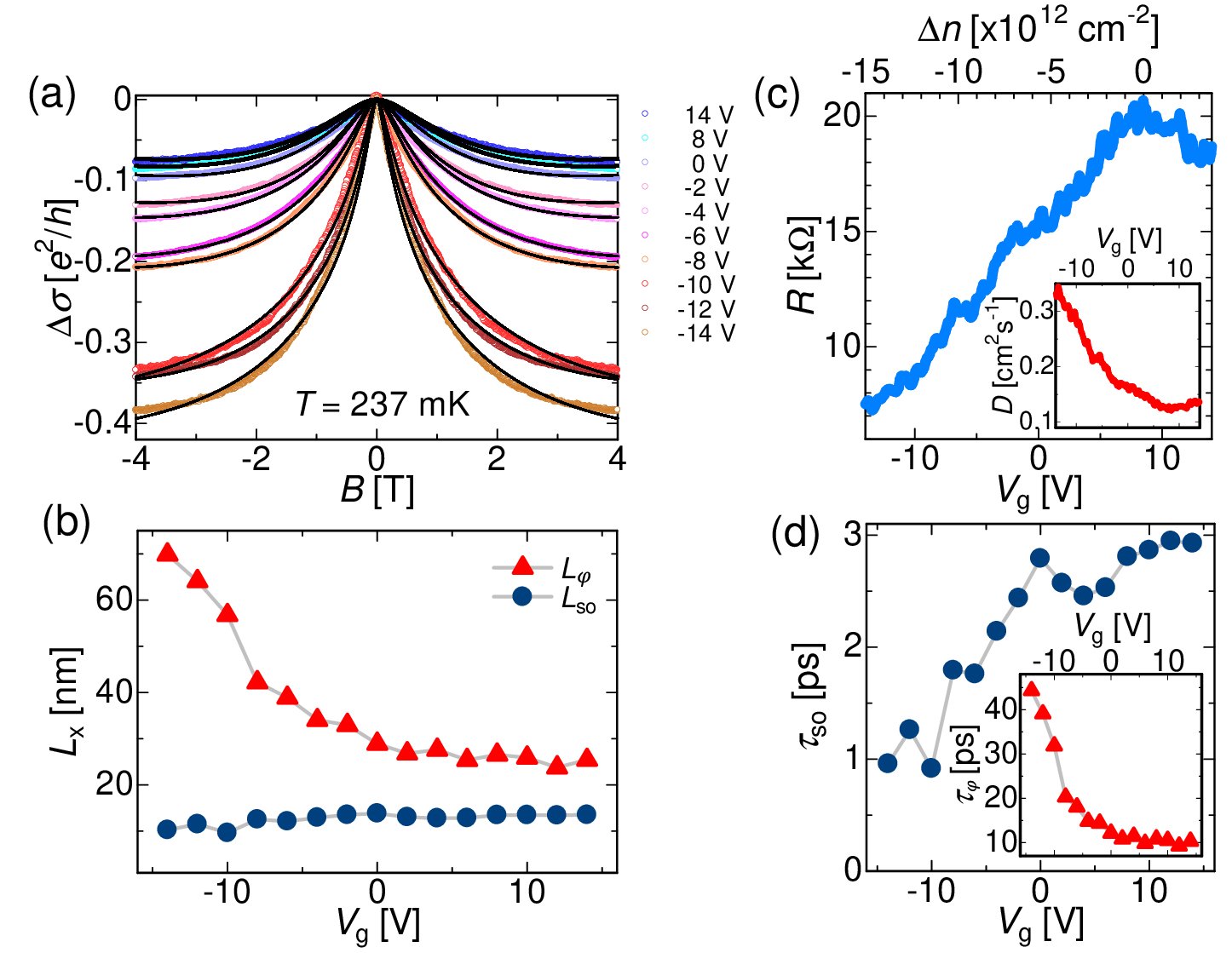}
\caption{Gate-dependent transport for the 1 L sample at 237 mK. (a) Magnetoconductivity variation $\Delta \sigma (B)$ as a function of $B$ measured at selected gate voltages ($V_g$). The solid lines are the fits using the equation (\ref{HLN}).  (b) $L_\varphi$ and $L_{\rm so}$ as a function of $V_g$. (c) $V_g$ dependence of the resistance $R$. Inset: The diffusion coefficient $D$ as a function of $V_g$ calculated from $R$ in the main panel. The top axis is in the unit of $\Delta n = C_g (V_g-V_0)/e$, where $C_g$ and $V_0$ are the capacitance of the hBN gate dielectric and the gate voltage at the charge neutrality point, respectively. (d) $V_g$ dependence of $\tau_{\rm so}$ calculated from $L_{\rm so}$ obtained from the fits in (a), exhibiting the threefold variation. The inset shows a similar $V_g$ dependence for $\tau_\varphi$.}
\label{fig3}
\end{center}
\end{figure*}

\textcolor{black}{In this subsection we show some representative WAL curves obtained from few-layer $T_{\rm d}$-MoTe$_2$ in different thicknesses. WAL arises from quantum interference of time-reversed electron trajectories in diffusive conductors with strong spin--orbit interaction (SOI), leading to a characteristic magnetoconductivity peak around zero magnetic field \cite{Bergmann_1984}. The WAL line shape is governed primarily by the phase coherence length $L_\varphi$ and the spin--orbit length $L_{\rm so}$. $L_\varphi$ characterizes the length scale over which electrons preserve their phase coherence, whereas $L_{\rm so}$ represents the characteristic length over which electron spins are randomized by spin--orbit scattering. By fitting the WAL curves, these characteristic lengths can be extracted quantitatively.} We first show in Fig.~\ref{fig1}(a) the magnetoconductivity variations, $\Delta \sigma(B) = \sigma(B)-\sigma(0)$, for three different thicknesses (1 L, 2 L, and 4 L). Here the conductivity is derived from the measured resistance $R$ assuming the two-dimensional system as $\sigma = d/(RW)$, where $d$ and $W$ are the center-to-center distance between the two electrodes and the width of the sample, respectively. All curves exhibit peaks around zero magnetic field, a hallmark of WAL, indicating the presence of strong SOIs in $T_{\rm d}$-MoTe$_2$. Figure~\ref{fig1}(b) displays WAL curves for a 6 L sample at different temperatures, where the peak around the zero magnetic field is suppressed with increasing temperature. These curves allow quantitative evaluation of the spin--orbit scattering time as well as the phase coherence time \cite{Bergmann_1984}.
For quantitative analysis, we use the theoretical framework developed by Iordanskii, Lyanda-Geller, and Pikus (ILP), which assumes a system with broken inversion symmetry and explicitly takes the spin--orbit field $B_{\rm so}$ into account \cite{Iordanskii_1994, Knap_1996, Niu_2020, Nakamura_2012}. \textcolor{black}{Here we employ a standard formula for WAL, and do not explicitly include contributions from topological channels such as Weyl fermions and topological surface or edge states. This is because, even if a Weyl semimetal phase persists in few-layer $T_{\rm d}$-MoTe$_2$, it has not been experimentally established that the Fermi level lies sufficiently close to the Weyl points for Weyl-related transport to make a significant contribution. Moreover, our magnetotransport measurements do not allow us to separately identify contributions from topological states.} In the ILP model, $\Delta \sigma (B)$ under a perpendicular magnetic field $B$ is expressed as
\begin{widetext}
\begin{equation}
\Delta \sigma(B)  =  \frac{Ne^2}{\pi h} \left[F \left(\frac{B_\varphi+B_{\rm so}}{B} \right) + \frac{1}{2} F \left(\frac{B_\varphi+2B_{\rm so}}{B} \right) 
- \frac{1}{2} F \left(\frac{B_\varphi}{B} \right) \right],
\label{HLN}
\end{equation}
\end{widetext}
where
\begin{equation}
F(z) = \psi \left( \frac{1}{2} + z \right) - \ln (z)
\end{equation}
with $\psi$ as the Digamma function and $B_{\{\varphi, \rm{so}\}}= \hbar/4eL_{\{\varphi, \rm{so}\}}^2$. 
Here $L_\varphi$ and $L_{\rm so}$ are the phase coherence and spin--orbit length, respectively. 
\textcolor{black}{They can be associated with the phase coherence time ($\tau_\varphi$) and spin--orbit time ($\tau_{\rm so}$) as $L_{\{\varphi, \rm{so} \}} = \sqrt{D \tau_{\{ \varphi, \rm{so} \}}}$ with the diffusion coefficient $D$.}
Theoretical fits for experimental results provide estimates of $L_\varphi$ and $L_{\rm so}$, both of which are treated as fitting parameters in Eq. (\ref{HLN}). 

\textcolor{black}{To account for the semimetallic nature of $T_{\rm d}$-MoTe$_2$, we introduce a prefactor $N$ in Eq. (\ref{HLN}), which is treated as the effective number of coherent transport channels \cite{Wang_2022}. The effective value of $N$ may depend on the relative mobilities and scattering rates of the individual channels. For each device, we compared fits obtained with $N$=1 and $N$=2 using the same fitting range and procedure. The value of $N$ was selected on a device-by-device basis according to the fit quality and was then kept fixed throughout the gate-voltage series for that device. In particular, $N$=1 was adopted for the 1 L device because it consistently yielded better fits than $N$=2 over the entire gate-voltage range. Therefore, the observed gate dependence, including the approximately threefold modulation of $\tau_{\rm so}$, does not arise from changing $N$ with gate voltage.}

The solid lines shown in Fig.~\ref{fig1}(a) are fits based on Eq. (\ref{HLN}). 
\textcolor{black}{The ILP model reproduces the essential low-field features of the experimental WAL curves and provides a reasonable description of the magnetoconductivity in the diffusive regime. Small systematic deviations emerge at higher magnetic fields, possibly due to contributions not captured by the approximate ILP model.}

Note that Eq.~(\ref{HLN}) is an approximated version of the full ILP formula, where the $k$-cubic SOI is considered while the $k$-linear SOI is not taken into account  \cite{Iordanskii_1994}. This equation is often employed for analyzing WAL signals in systems lack of inversion symmetry \cite{Nakamura_2012, Niu_2020}. \textcolor{black}{In Appendix A, we assess the validity of Eq.~(\ref{HLN}) by fiting the experimental results using the full ILP formula while varying the ratio between the $k$-cubic and $k$-linear SOI terms for different layer thicknesses and for a monolayer sample under different vertical electric fields. The fitting results systematically favor the small $k$-linear SOI regime, supporting the use of Eq. (\ref{HLN}) in the systematic WAL analysis presented below. Detailed fitting comparisons and quantitative evaluations of the fitting quality are provided in Appendix A.} 
\begin{figure*}[!tb]
\begin{center}
\includegraphics[width=14cm,clip]{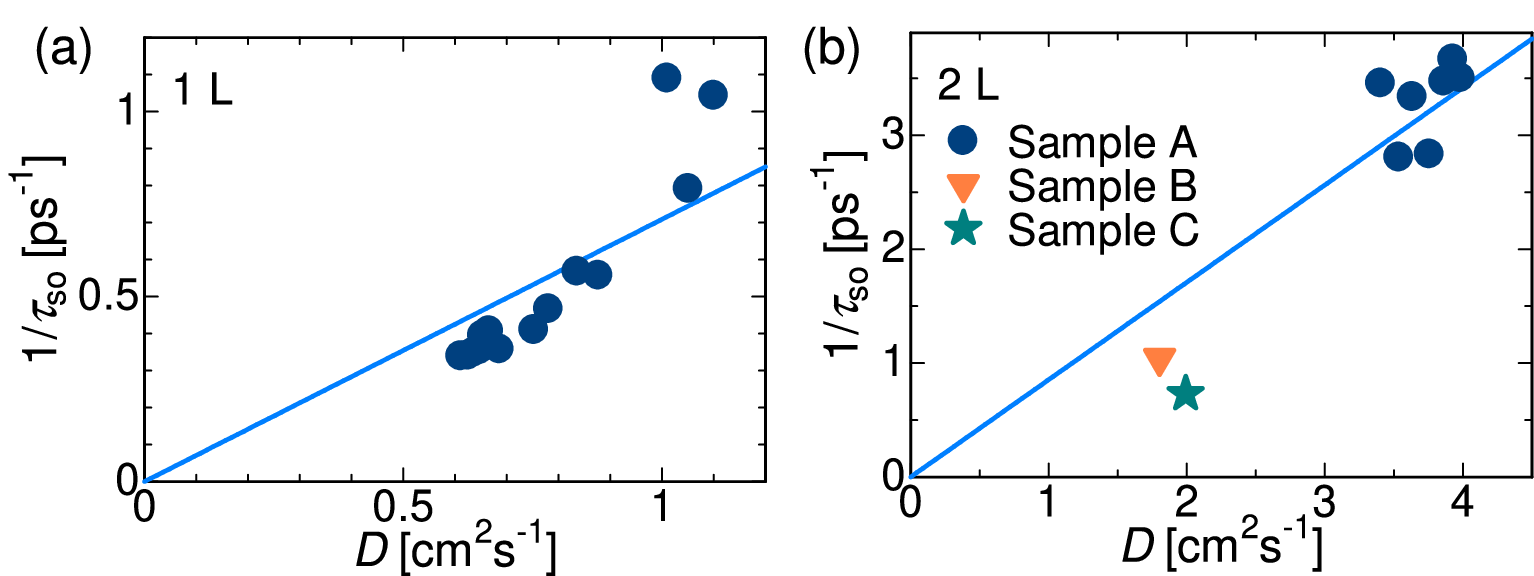}
\caption{(a) Inverse of $\tau_{\rm so}$ as a function of the diffusion coefficient $D$ for a single 1 L $T_{\rm d}$-MoTe$_2$ sample. An almost linear relation between the two quantities shows the dominant DP spin relaxation mechanism, as guided by the light-blue line proportional to $D$. (b) A similar plot as (a) for multiple 2 L samples. The light-blue line is proportional to $D$.}
\label{fig5}
\end{center}
\end{figure*}

From the fits, we extract $L_\varphi$ and $L_{\rm so}$.  
Figure~\ref{fig1}(c) shows the temperature dependence of $\tau_\varphi^{-1}$ for the 6 L sample, calculated from $L_\varphi$ and $D$ obtained from the resistance measurements at each temperature. Almost linear relation shows that the phase relaxation is driven by electron--electron scattering \cite{Altshuler_1982}. We present in Fig.~\ref{fig1}(d) the spin--orbit length $L_{\rm so}$ as a function of thickness. $L_{\rm so}$ is in between 6--45 nm, and these values are in accordance with theoretical estimates \cite{Vila_2021, Xu_2024} and also experimentally obtained value in a recent experiment for a similar LS-TMD WTe$_2$ \cite{Zhang_2017}. 



\subsection{Gate modulation of spin transport in monolayer $T_{\rm d}$-MoTe$_2$}
Next, we study the gate modulation of spin transport in $T_{\rm d}$-MoTe$_2$. In this subsection, we focus on effects from the carrier density modulation by gating. In bulk crystals, the semimetallic nature leads to strong screening, making it difficult to modulate spin and charge transport using conventional electrostatic gating. However, when the sample thickness becomes comparable to or smaller than the nanometer-scale screening length, a gate electric field can effectively penetrate the material and control transport. Figure~\ref{fig3}(a) shows $\Delta \sigma (B)$ measured in a 1 L sample at different $V_g$ at 237 mK. Note that this sample does not show superconductivity at this temperature \cite{Wakamura_2026}. \textcolor{black}{The pronounced change in $\Delta \sigma (B)$ with $V_g$ indicates that the spin-relaxation properties are significantly modulated by electrostatic gating.}

By fitting the magnetoconductivity curves [solid lines in Fig.~\ref{fig3}(a)], we extract $L_\varphi$ and $L_{\rm so}$, which are summarized in Fig.~\ref{fig3}(b). While $L_\varphi$ decreases monotonically with increasing $V_g$, $L_{\rm so}$ exhibits only a weak gate dependence. We can obtain $\tau_{\{\varphi, {\rm so}\}}$ from $L_{\{\varphi, {\rm so}\}}$ using the diffusion coefficient $D$. $D$ is calculated from the conductivity $\sigma$ as $D = h \sqrt{\sigma h}/(4 m e \pi \sqrt{2})$, where $m = 0.37 m_0$ is the effective mass with $m_0$ the electron mass \cite{Rhodes_2021}. The longitudinal resistance, shown in Fig.~\ref{fig3}(c), exhibits a strong gate dependence with a charge neutrality point at $V_g \sim$ 8 V, indicating that the sample is hole-doped at $V_g = 0$ V. The $V_g$ dependence of $D$ is presented in the inset of Fig.~\ref{fig3}(c). Using this $D$, we calculate and plot $\tau_{\rm so}$ as a function of $V_g$ in Fig.~\ref{fig3}(d). $\tau_{\rm so}$ increases by a factor of three as with $V_g$ varies from $-14$ to 14 V. This pronounced gate modulation of $\tau_{\rm so}$ constitutes a central result of this work. Previous studies of WAL in thin $T_{\rm d}$-MoTe$_2$ have either not explored gate modulation \cite{Naylor_2016, Wang_2018} or reported only weak modulation, likely due to limited electrostatic tunability \cite{Song_2018_WAL}. \textcolor{black}{Note that this threefold modulation of $\tau_{\rm so}$ does not imply a corresponding threefold modulation of the intrinsic SOI because $\tau_{\rm so}$ also depends on momentum relaxation, as discussed in the later subsection.}




We also investigate the origin of the gate dependence of $L_\varphi$. From Figs.~\ref{fig3}(b) and (c), we find that $L_\varphi$ decreases as the resistance increases. This can be explained by enhanced electron--electron scattering, as the screening becomes weaker when the carrier density decreases. It is also consistent with the relation between $\tau_\varphi$ and $V_g$ as shown in the inset of Fig.~\ref{fig3}(d), where $\tau_\varphi$ becomes smaller with increasing resistance. At this low temperature ($T$ = 237 mK), phonons are negligible, thus electron--phonon scattering is ineffective. \textcolor{black}{We note the electron--electron and momentum scattering play different roles. At low temperatures, electron--electron interactions can provide a dominant phase relaxation mechanism, whereas momentum scattering, including impurity scattering, affects the spin relaxation through the DP mechanism.}

\subsection{Identification of spin relaxation mechanisms}
The opposite trend for $\tau_{\rm so}$ and $D$ in regard to $V_g$, shown in Figs.~\ref{fig3}, can be understood from the underlying spin relaxation mechanism. Here we investigate the dominant spin relaxation mechanism in few-layer $T_{\rm d}$-MoTe$_2$. In diffusive metals such as $T_{\rm d}$-MoTe$_2$, two spin relaxation mechanisms are generally considered \cite{Fabian_1999}: the Elliott--Yafet (EY) mechanism \cite{Elliot_1954} and the D'yakonov--Perel' (DP) mechanism \cite{Dyakonov_1971}. In the EY mechanism, electron spins relax via momentum scattering through spin--orbit coupling, leading to the relation $\tau_{\rm so} \propto \tau_p$, where $\tau_p$ is the momentum relaxation time. On the other hand, the DP mechanism is closely associated with systems lacking inversion symmetry. Spin splitting caused by broken inversion symmetry can be interpreted as a momentum-dependent effective Zeeman field $\mathbf{B}(\mathbf{k})$. Between successive momentum scattering events, electron spins precess around $\mathbf{B}(\mathbf{k})$, which promotes spin relaxation. In this case, the spin relaxation time follows the inverse dependence $\tau_{\rm so} \propto \tau_p^{-1}$, in contrast to the EY mechanism.

The dominant spin relaxation mechanism can therefore be identified by examining the relationship between $\tau_{\rm so}$ and $D$ ($\propto \tau_p$). We investigate the dependence of $\tau_{\rm so}^{-1}$ on $D$ for 1 L and 2 L $T_{\rm d}$-MoTe$_2$ samples. Figure~\ref{fig5}(a) shows the results for the 1 L sample. The large gate tunability of $D$ in the 1 L sample allows us to determine the dominant spin relaxation mechanism within a single device. We find that $\tau_{\rm so}^{-1}$ is approximately proportional to $D$, demonstrating that the DP mechanism governs spin relaxation in 1 L $T_{\rm d}$-MoTe$_2$. We also show the corresponding results for the 2 L samples in Fig.~\ref{fig5}(b). Due to the limited gate tunability of $D$ within a single device in 2 L sample, data from multiple devices are plotted. The observed increase of $\tau_{\rm so}^{-1}$ with increasing $D$ indicates that the DP mechanism dominates spin relaxation in the 2 L samples. 

The dominance of the DP mechanism in 2 L samples is consistent with the absence of inversion symmetry in 2 L $T_{\rm d}$-MoTe$_2$, as in the bulk. By contrast, inversion symmetry is preserved in 1 L $T_{\rm d}$-MoTe$_2$, whose crystal structure is the same as inversion-symmetric 1$T'$-MoTe$_2$. However, inversion symmetry can be broken by the presence of a substrate and a gate electric field normal to the sample plane \cite{Zhang_2018}. In addition, although our samples are encapsulated with hBN on both sides, asymmetry between the top and bottom interfaces may be introduced during device fabrication, for example, through surface oxidation as reported for WTe$_2$ \cite{Fei_2018}. A difference in the dielectric environment between the two sides can also break inversion symmetry \cite{Wu_2013}. Another possible factor is an intermediate metastable state between the inversion-symmetric 1$T'$ phase and the inversion-asymmetric $T_{\rm d}$ phase reported by a recent Raman spectroscopy study \cite{Huang_2019}. Such extrinsic symmetry breaking can generate an effective spin--orbit field, thereby enhancing DP-type spin relaxation relative to the EY mechanism \cite{Xu_2024}.

\begin{figure*}[!tb]
\begin{center}
\includegraphics[width=17cm,clip]{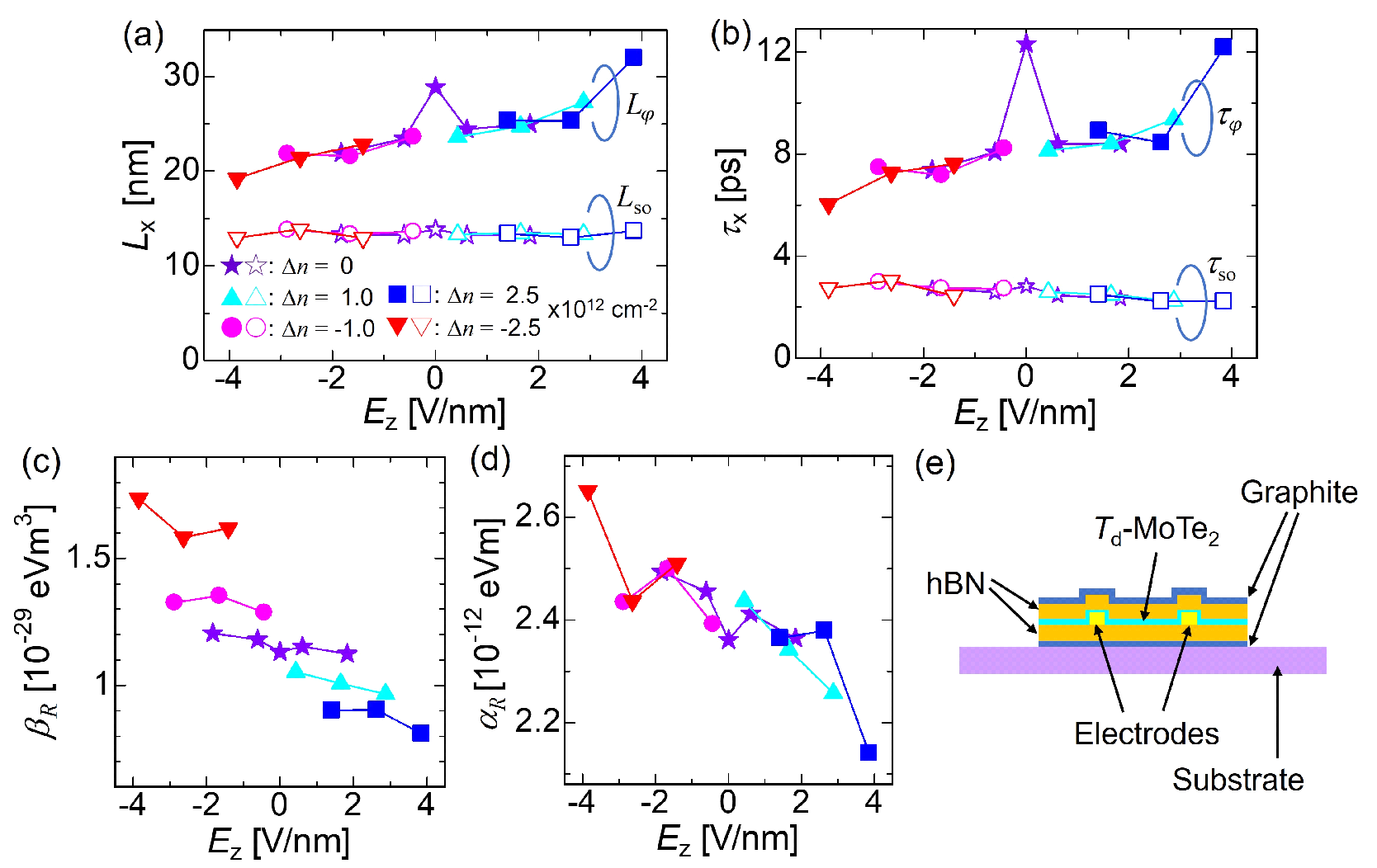}
\caption{(a) Vertical electric field ($E_z$) dependence of the characteristic length ($L_\varphi$ and $L_{\rm so}$) at different carrier densities ($\Delta n$) for the 1 L sample. (b) $\tau_\varphi$ and $\tau_{\rm so}$, (c) $\beta_R$ and (d) $\alpha_R$ for different $\Delta n$ plotted as a function of $E_z$. Correspondence between the symbol and $\Delta n$ is the same for three figures. All data are taken at $T$ = 237 mK. (e) Schematic illustration of the dual-gate device structure.}
\label{fig4}
\end{center}
\end{figure*}

\subsection{Vertical electric field dependence of spin relaxation}

To disentangle the effects of vertical electric fields from the carrier density modulation, we performed measurements on dual-gate samples incorporating both top and bottom gates. The vertical electric field $E_z$ is given by
$E_z = \varepsilon_t V_t / d_t - \varepsilon_b V_b / d_b$, where $\varepsilon_{t,b}$, $V_{t,b}$, and $d_{t,b}$ denote the relative dielectric constant, gate voltage, and thickness of the hBN layers for the top ($t$) and bottom ($b$) gates, respectively.

Figure~\ref{fig4}(a) shows the dependence of $L_\varphi$ and $L_{\rm so}$ on $E_z$ for the 1 L sample measured at 237 mK [see the device structure shown in Fig.~\ref{fig4}(e)]. To vary $E_z$ over a wide range, measurements were performed for several values of $\Delta n$ [$=\varepsilon_0 \varepsilon_t V_t/(e d_t) + \varepsilon_0 \varepsilon_b V_b/(e d_b)$, where $\varepsilon_0$ and $e$ are the vacuum dielectric constant and the elementary charge, respectively]. Note that the range over which $\Delta n$ is varied here is much smaller than in Figs.~\ref{fig3} [see the top axis of Fig.~\ref{fig3}(c)], indicating that the observed behavior primarily reflects the effect of the vertical electric field rather than carrier doping. While $L_\varphi$ shows a clear dependence on $E_z$, $L_{\rm so}$ remains nearly constant.
In Fig.~\ref{fig4}(b), we also plot $\tau_{\{\varphi,\rm so\}}$ derived from $L_{\{\varphi,\rm so\}}$. The $E_z$ dependence is more pronounced for $\tau_\varphi$ than $\tau_{\rm so}$. Although the variation is modest, $\tau_{\rm so}$ shows a systematic decrease with increasing $E_z$. This trend is qualitatively consistent with a previous theoretical study on the conduction bands of WTe$_2$ \cite{Xu_2024}. The extracted $L_{\{\varphi,\rm so\}}$ and $\tau_{\{\varphi,\rm so\}}$ are not symmetric with respect to $E_z = 0$, suggesting the presence of an internal electric field, as discussed above.


To quantitatively evaluate the effect of $E_z$ on the spin--orbit interaction, we first estimate the cubic Rashba parameter $\beta_R = \Delta_{\rm so}/k_F^3$, where $k_F = \sqrt{2\pi n}$ is the Fermi wave number and $n$ is the total carrier density obtained from magnetoresistance fitting \cite{Wakamura_2026}, considering that Eq. (\ref{HLN}) assumes the cubic Rashba SOI. Based on the DP spin relaxation, the spin splitting is given by $\Delta_{\rm so} = \hbar/(2\sqrt{\tau_p \tau_{\rm so}})$ \cite{Minkov_2005}, and the resulting $\beta_R$ as a function of $E_z$ is shown in Fig.~\ref{fig4}(c). \textcolor{black}{Note that $\beta_R$ (also $\alpha_R$ as discussed below) depends not only on $E_z$ but also on $n$, through the $k_F = \sqrt{2\pi n}$ and $\tau_p$.} \textcolor{black}{The carrier doping effect is manifested as vertical shifts in $\beta_R$ for different $\Delta n$ due to the non-negligible change of $k_F^3$ in the denominator. On the other hand, although the modulation of $\beta_R$ by $E_z$ is moderate, it is clearly visible at fixed $\Delta n$ when comparing the data points of the same color in Fig.~\ref{fig4}(c).} We also plot the linear Rashba parameter calculated as $\alpha_R = \Delta_{\rm so}/k_F$ in Fig.~\ref{fig4}(d), which presents more than 20 $\%$ variation in the same range of $\Delta n$. As for the Fermi level dependence of the Rashba parameters, both Rashba parameters increase as the hole density decreases (from positive to negative $\Delta n$), corresponding to the Fermi level approaching the valence band edge. This behavior is consistent with previous studies on the related material monolayer WTe$_2$, where electric-field-induced spin splitting becomes more pronounced near the band edge \cite{Xu_2018, Xu_2024}.




\section{Conclusion}

In conclusion, we have characterized spin relaxation and demonstrated electrostatic control of spin transport in few-layer $T_{\rm d}$-MoTe$_2$. Pronounced WAL signals reveal strong SOIs, and analysis based on the ILP model yields spin--orbit lengths on the order of several tens of nanometers, in quantitative agreement with theoretical estimates. In 1 L $T_{\rm d}$-MoTe$_2$, we achieve substantial gate modulation of the spin--orbit time, reaching a threefold variation. \textcolor{black}{The vertical electric field dependence further demonstrates electrical control of Rashba parameters}. We also identify the dominant spin relaxation mechanism, showing that DP relaxation governs spin dynamics in both 1 L and 2 L samples. 

While previous experiments have reported significantly longer spin relaxation lengths \cite{Song_2020}, the origin of this discrepancy remains an open question and may reflect differences in measurement schemes. Overall, this work establishes LS-TMDs as a platform for electrically tunable spin transport and advances the understanding of spin relaxation in these materials.

We gratefully acknowledge J. Nitta and H. Yang for fruitful discussions, and thank M. Imai, S. Sasaki, and H. Murofushi for their support in the experiments.
This project is financially supported in part by the JPSJ KAKENHI (Grant No. JP23H04869, No. JP21H01022, No. JP21H01041, No. JP21H04652, No. JP21K18181, No. JP21H05236, No. JP20H00354, No. JP19H05790, No. JP22H00112, No.  JP24H00827 and No. JP26K00669). K.W. and T.T. acknowledge support from JSPS KAKENHI (Grants No. 21H05233 and No. 23H02052), the CREST (JPMJCR24A5), JST, and the World Premier International Research Center Initiative (WPI), MEXT, Japan.


\appendix

\section{Fitting the experimental results using the full Iordanskii Lyanda-Geller Pikus (ILP) formula}


While in the main text we employ an approximated version of the ILP formula, in this subsection we compare it with the full ILP formula \cite{Iordanskii_1994}. The full ILP formula is expressed as
\begin{widetext}
\begin{multline}
\Delta \sigma = -\frac{Ne^2}{4 \pi^2 \hbar} \bigg\{ \frac{1}{a_0} + \frac{2a_0 + 1 + (\widetilde{B}_{\rm so}/B)}{a_1[a_0 + (\widetilde{B}_{\rm so}/B)]-2 (B_{\rm so1}/B)} + 2 \ln \bigg( \frac{B_{tr}}{B} \bigg) + \psi \bigg( \frac{1}{2} + \frac{B_\varphi}{B} \bigg) + 3C \\
- \sum_{n=1}^\infty \bigg[ \frac{3}{n}- \frac{3a_n^2 + 2a_n (\widetilde{B}_{\rm so}/B) - 1 -2(2n+1)(B_{\rm so1}/B)}{[a_n + (\widetilde{B}_{\rm so}/B)] a_{n-1} a_{n+1} -2 (B_{\rm so1}/B)[(2n+1)a_n-1]} \bigg] \bigg\},
\label{ILP}
\end{multline}
\end{widetext}
where 
\begin{widetext}
\begin{equation}
\begin{aligned}
\widetilde{B}_{\rm so} = B_{\rm so1} + B_{\rm so3}, \quad B_{\rm so1} = \frac{\hbar}{4eD} |\Omega_1|^2 \tau_1, \quad B_{\rm so3} = \frac{\hbar}{4eD} |\Omega_3|^2 \tau_3 = B_{\rm so}, \quad B_{tr}=\hbar/(4eD\tau_p), \\
\frac{1}{\tau_m} = \int [1 - \cos(m\theta)] W(\theta) d\theta, \quad a_n = n + 1/2 + (B_\varphi/B) + (\widetilde{B}_{\rm so}/B).
\label{params}
\end{aligned}
\end{equation}
\end{widetext}
$W(\theta)$ and $C$ are the scattering probability by an angle $\theta $ and the Euler constant, respectively. $\Omega_1$ and $\Omega_3$ are the spin precession vector originated from the Rashba and Dresselhaus SOI, and proportional to the wave number $k$ and $k^3$, respectively. $B_{\rm so3}$ is equivalent to $B_{\rm so}$ in the main text. Note that, for simplicity, we set $\tau_1 = \tau_3 = \tau_p$. To fit the experimental data, we truncate the infinite sum in (\ref{ILP}) at $n = 20000$. The parameter $B_{tr}$ is determined independently from transport measurements. The equation (\ref{HLN}) in the main text is obtained by setting $B_{\rm so1} = 0$ in Eqs. (\ref{ILP}) and (\ref{params}) \cite{Iordanskii_1994}.



\begin{figure*}[!tb]
\begin{center}
\includegraphics[width=17cm,clip]{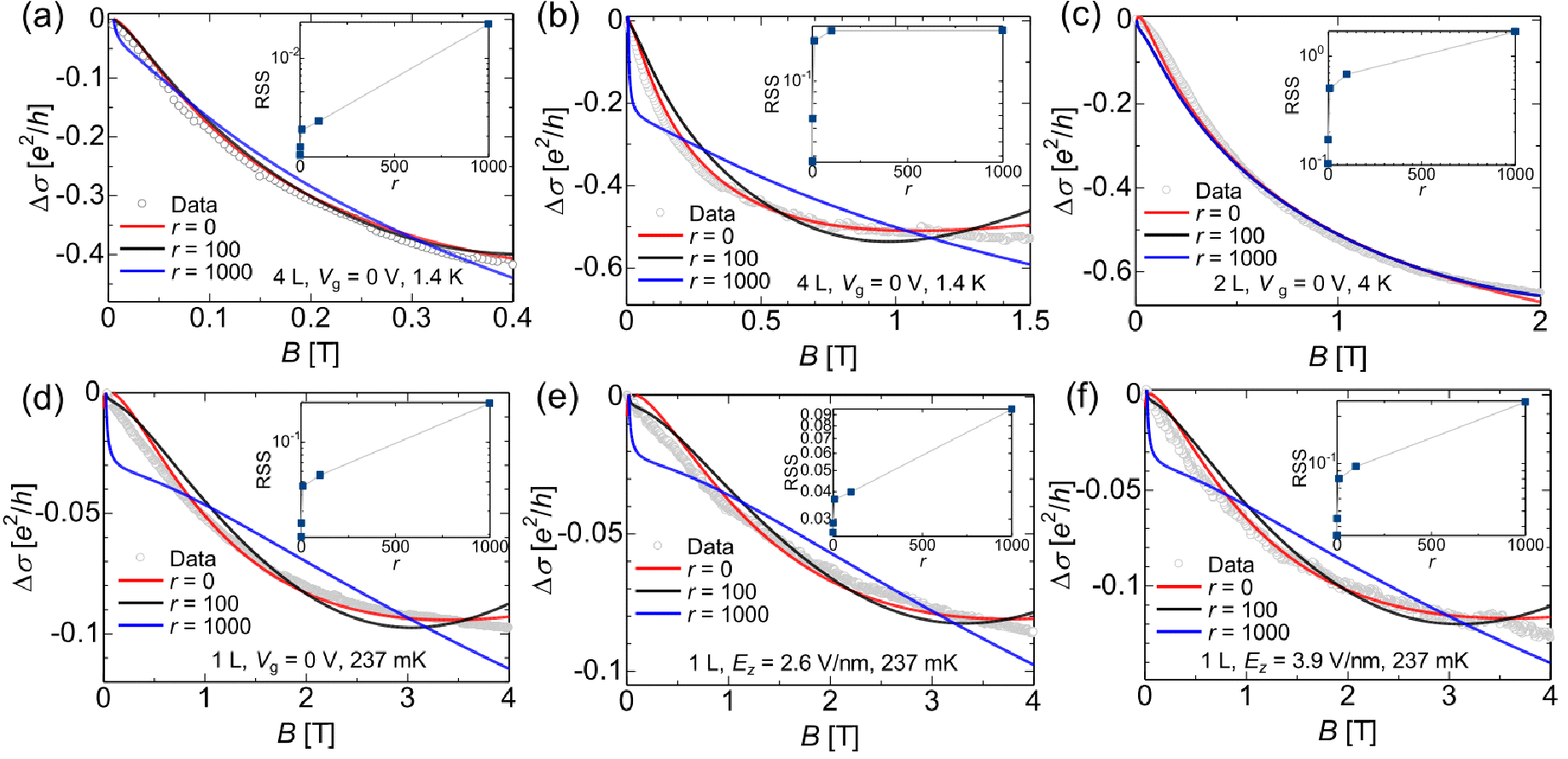}
\caption{Comparison of the WAL fits using the Eq. (\ref{ILP}) for different $r$. The inset shows the evolution of RSS as a function of $r$. \textcolor{black}{(a) and (b) are the data taken from a 4 L sample (MTS3) at 1.4 K and $V_g$=0 for the different fitting range. (c) is from a 2 L (MTS4) sample, and (d)-(f) compares the results for a 1 L (MTS11) samples under different vertical electric field ($E_z$) conditions.} }
\label{figILP}
\end{center}
\end{figure*}





\begin{table*}
\caption{Values of the fitting parameters obtained from the fits using Eq. (\ref{ILP}) for different $r$ \textcolor{black}{for the case shown in Fig~\ref{figILP}(a).}}
\label{table3}
\begin{ruledtabular}
\begin{tabular}{rcccc}
$r$ & $B_\varphi$ [T] & $B_{\rm so1}$ [T] & $B_{\rm so3}$ [T] & RSS \\ \hline
0 & $1.30 \times 10^{-2} \pm 6.66 \times 10^{-5}$ & 0 & $0.164 \pm 2.37 \times 10^{-3}$ & $1.35 \times 10^{-3}$ \\
1 & $1.34 \times 10^{-2} \pm 6.99 \times 10^{-5}$ & $0.127 \pm 1.74 \times 10^{-3}$ & $0.127 \pm 1.74 \times 10^{-3}$ & $1.60 \times 10^{-3}$ \\
10 & $1.41 \times 10^{-2} \pm 7.61 \times 10^{-5}$ & $0.311 \pm 3.75 \times 10^{-3}$ & $3.11 \times 10^{-2} \pm 3.75 \times 10^{-4}$ & $2.29 \times 10^{-3}$ \\
100 & $1.43 \times 10^{-2} \pm 7.93 \times 10^{-5}$ & $0.345 \pm 3.98 \times 10^{-3}$ & $3.45 \times 10^{-3} \pm 3.98 \times 10^{-5}$ & $2.73 \times 10^{-3}$ \\
1000 & $2.14 \times 10^{-2} \pm 6.88 \times 10^{-4}$ & $4.46 \pm 0.486$ & $4.46 \times 10^{-3} \pm 4.86 \times 10^{-4}$ & $2.02 \times 10^{-2}$ \\
\end{tabular}
\end{ruledtabular}
\end{table*}
To investigate effects of $B_{\rm so1}$ and $B_{\rm so3}$ terms in the fits, we define the factor $r$ given by the ratio between the two terms ($r=B_{\rm so1}/B_{\rm so3}$), and at fixed $r$, we carry out fitting by treating $B_\varphi$ and $\widetilde{B}_{\rm so}$ as fitting parameters. Figure~\ref{figILP} presents the fits at different $r$ from $r=0$ to $r=1000$ \textcolor{black}{for the samples with different thicknesses and vertical electric fields}. The obtained fitting parameters and the residual sum of squares (RSS) \textcolor{black}{for the condition shown in Fig.~\ref{figILP}(a)} are given in Table \ref{table3} \textcolor{black}{as representative results}. The inset of Fig.~\ref{figILP} shows the RSS as a function of $r$, which clearly shows that the better fit is obtained in the limit of $r \rightarrow 0$ ($B_{\rm so1} \rightarrow$ 0) and validates the analysis using Eq. (\ref{HLN}).

\section{Inplane magnetic field dependence of weak antilocalization}
Previous theoretical and experimental studies for few-layer $T_{\rm d}$-MoTe$_2$ have pointed out the spin polarization anisotropy, particularly for the conduction bands \cite{Vila_2021, Cui_2019, Rhodes_2021, Li_2024, Wang_Yang_2018}. To explore effects of the spin polarization anisotropy on WAL in few-layer $T_{\rm d}$-MoTe$_2$, we measure the inplane magnetic-field dependence of WAL using a 4 L $T_{\rm d}$-MoTe$_2$ sample. 
\begin{figure}[tb!]
\begin{center}
\includegraphics[width=8cm,clip]{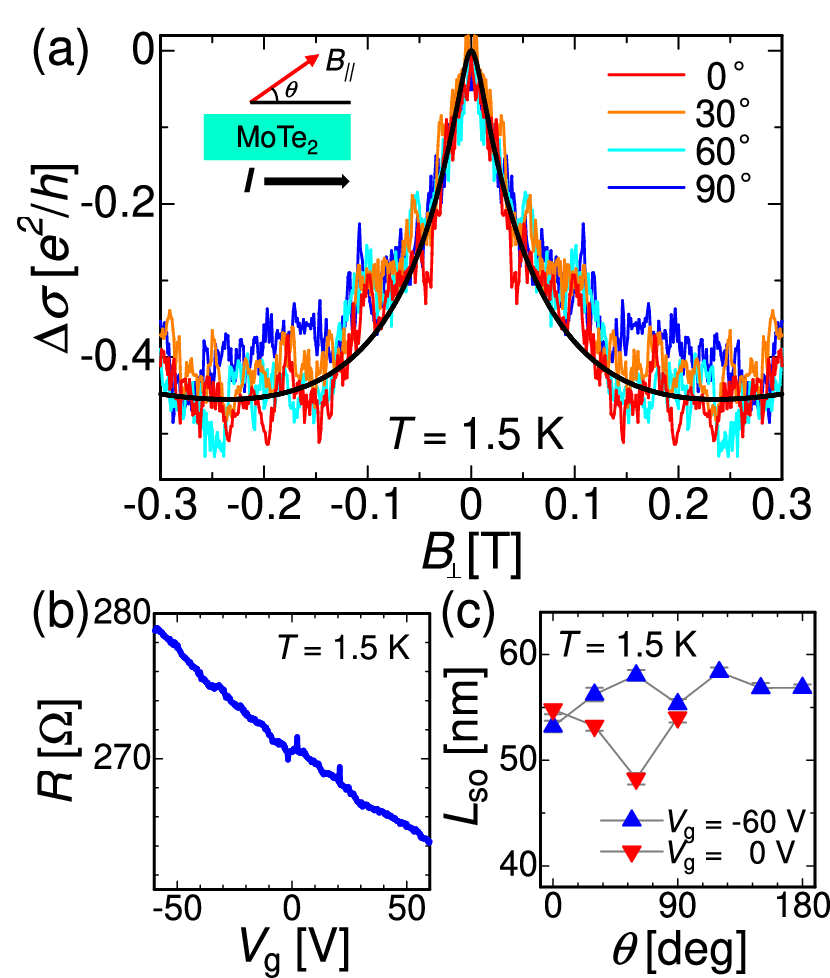}
\caption{(a) $\Delta \sigma (B)$ curves taken from a 4 L sample \textcolor{black}{(MTS22)} with in-plane magnetic field ($B_\parallel$ = 1 T) at four different angles and at $V_g$ = 0 V. The angle is defined in the inset. The black solid curve represents the fit using the ILP formula for the data at $\theta = 0^\circ$. (b) Gate voltage dependence of the resistance of the sample. It is electron-doped at $V_g$ = 0 V and the Fermi level approaches to the CNP as $V_g$ is swept toward a negative voltage. (c) Full angular dependence of $L_{\rm so}$ at $V_g$ = $-60$ V, shown together with the data at $V_g$ = 0 V. Particularly for $V_g$ = $-60$ V, $L_{\rm so}$ is almost independent of the inplane magnetic field of $B_\parallel$ = 1 T.}
\label{fig_inplane}
\end{center}
\end{figure} 

Figure~\ref{fig_inplane}(a) presents $\Delta \sigma(B_\perp)$ measured at 1.5 K and $V_g = 0$ V for different orientations of the in-plane magnetic field ($B_\parallel$). Here we explicitly write a perpendicular magnetic field as $B_\perp$ to distinguish it from $B_\parallel$. The magnitude of $B_\parallel$ is fixed at 1 T, and the relative angle $\theta$ between $B_\parallel$ and the longitudinal direction of the $T_{\rm d}$-MoTe$_2$ flake is varied (see the schematic in the inset of Fig.~\ref{fig_inplane} for the definition of $\theta$). The variation in the WAL curves is small, and $L_{\rm so}$ values obtained from the ILP fits are given in Fig.~\ref{fig_inplane}(c) for $\theta$ from $0^\circ$ to $90^\circ$. These results indicate that the spin polarization of the carriers is almost insensitive to an in-plane magnetic field of at least 1 T. 

We next measure the in-plane magnetic-field dependence of WAL in a different carrier-doping regime. In TMDs, the spin polarization structure changes depending on the position of the Fermi level, and the spin polarization anisotropy in the conduction bands becomes maximal when the Fermi level approaches the valley extrema. As shown in Fig.~\ref{fig_inplane}(b), the sample is electron-doped, and by applying a negative gate voltage, the Fermi level can be tuned closer to the charge neutrality point. The spin--orbit length $L_{\rm so}$, obtained from fits to the WAL signals at $V_g = -60$ V, is plotted in Fig.~\ref{fig_inplane}(c) as a function of $\theta$ over a broader range of $\theta = 0^\circ-180^\circ$. Similar to the case at $V_g = 0$ V, $L_{\rm so}$ is almost independent of the direction of $B_\parallel$.

Within the framework of the D'yakonov--Perel' spin relaxation mechanism (see the discussion in the main text for details), the spin-splitting energy $\Delta_{\rm so}$ is given by $\Delta_{\rm so} = \hbar / 2 \sqrt{\tau_p \tau_{\rm so}}$ \cite{Minkov_2005},
where $\tau_p$ is the momentum relaxation time. We estimate $\Delta_{\rm so} \approx 8.7$ meV, which is much larger than the energy scale of a 1 T magnetic field ($\sim 0.1$ meV), assuming a commonly used $g$-factor of $g = 2$. Therefore, the spins are so strongly locked by the spin--orbit field along the direction around $b$-axis that they remain unaffected by an in-plane magnetic field of $B_\parallel = 1$ T. To investigate the inplane spin polarization anisotropy, therefore, much larger $B_\parallel$ is required, which is inaccessible in our present experimental setup.
 
\section{Crystal structure of $T_{\rm d}$-MoTe$_2$}
\begin{figure}[tb!]
\begin{center}
\includegraphics[width=8cm,clip]{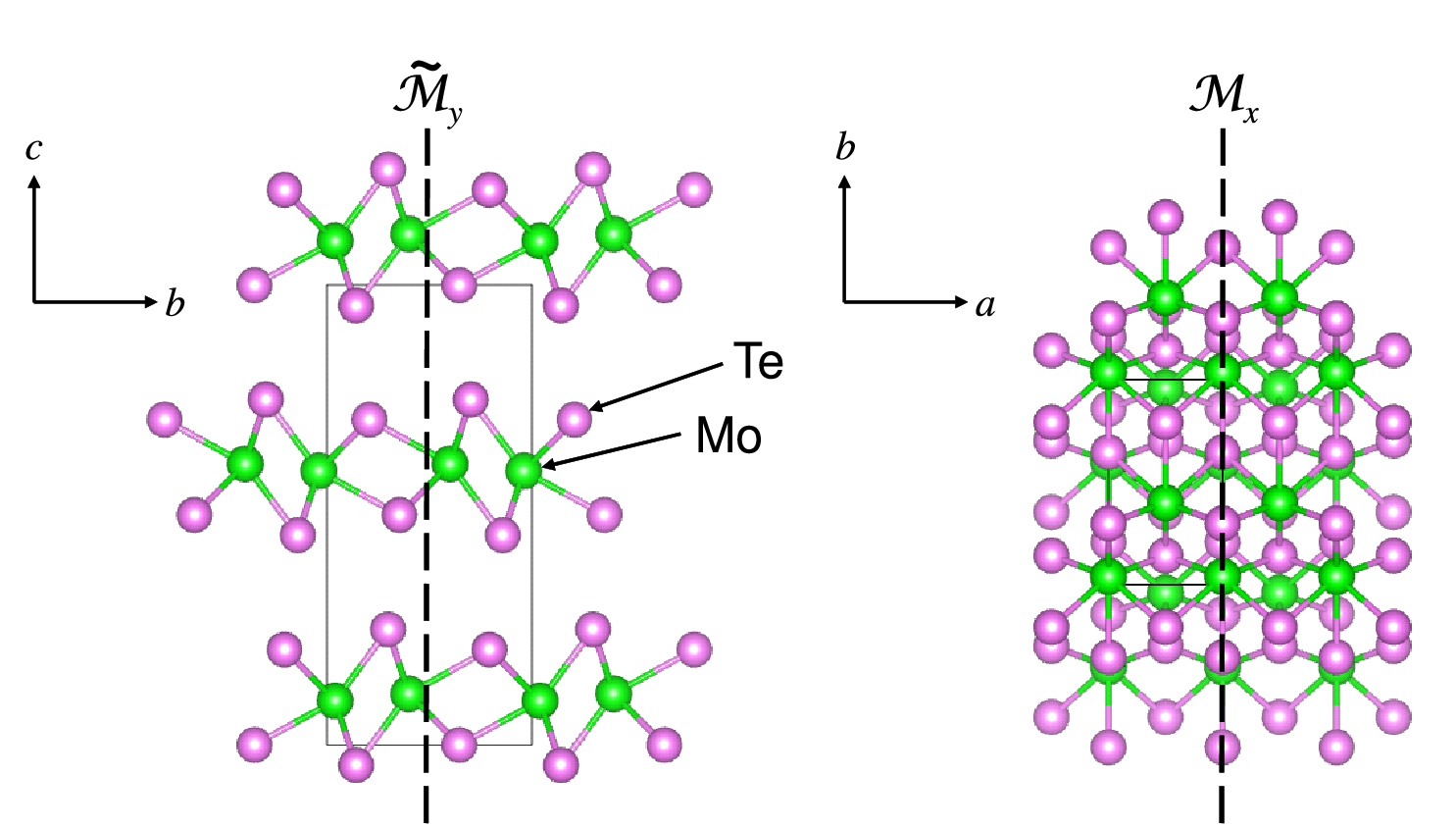}
\caption{The crystal structure of $T_{\rm d}$-MoTe$_2$ from the side view (left panel) and the top view (right panel). $\widetilde{\mathcal{M}}_y$ is the glide plane parallel to the $c$-axis and $\mathcal{M}_x$ is the mirror plane normal to the $a$-axis.}
\label{fig_crystal}
\end{center}
\end{figure}

We here show the crystal structure of $T_{\rm d}$-MoTe$_2$. Figures~\ref{fig_crystal} show the side (left panel) and top (right panel) view of the crystal structure of $T_{\rm d}$-MoTe$_2$. One mirror plane $\mathcal{M}_x$ normal to the $a$-axis as well as a glide plane $\widetilde{\mathcal{M}}_y$ exist in bulk $T_{\rm d}$-MoTe$_2$, the latter is absent for few-layer samples, presenting a highly anisotropic nature of the material. 

\begin{figure}[tb!]
\begin{center}
\includegraphics[width=8cm,clip]{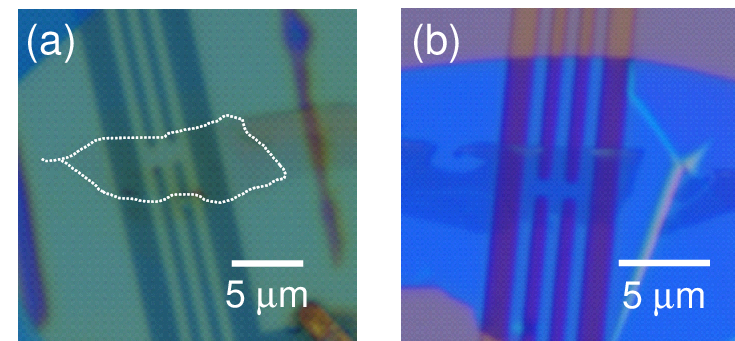}
\caption{Pictures of some of the samples used for this study. (a) presents a 2 L sample (MTS10) while (b) shows a 4 L sample (MTS22).}
\label{fig_samples}
\end{center}
\end{figure}

\textcolor{black}{\section{Summary of the samples and samples pictures}
In this subsection, we summarize the samples employed in this study with $B_{tr}$, a magnetic field range used for the fits, $L_\varphi$ and $L_{\rm so}$ obtained from the fits shown with the standard deviation. These data are presented in Table II. We also show the sample pictures of some of the representative samples for the present study in Fig.~\ref{fig_samples}.}

\begin{table*}[!tb]
\caption{\textcolor{black}{Summary of the samples employed in this study and parameters relevant to the fits.}}
\label{Table2}
\begin{ruledtabular}
\begin{tabular}{ccccccc}
 Sample Name & Number of Layers & $B_{tr}$ [T] & Fitting range [T] & $L_\varphi$ [nm] & $L_{\rm so}$ [nm] & $T$ [K] \\ \hline
 MTS11 & 1 & 7.8 $\times 10^2$ & 4 & 28.92 $\pm$ 0.03 & 13.85 $\pm$ 0.02 & 0.24   \\
 MTS4 & 2 & 4.7 -- 12 & 2 & 72.2 $\pm$ 0.1 & 10.9 $\pm$ 0.1 & 4  \\
 MTS10 & 2 & 4.5 & 2 & 25.0 $\pm$ 0.02 & 13.2 $\pm$ 0.01 & 4 \\
 MTS18 & 2 & 30 & 2 & 32.7 $\pm$ 0.1 & 18.8 $\pm$ 0.1 & 4 \\
 MTS3 & 4 & 2.5 & 2 & 43.6 $\pm$ 0.1 & 21.3 $\pm$ 0.1  & 4 \\
 MTS7 & 4 & 2.0 & 1.4 & 51.6 $\pm$ 0.2 & 27.3 $\pm$ 0.1 & 2.2 \\
 MTS22 & 4 & 1.64 & 0.5 & 158 $\pm$ 1.2 & 34.0 $\pm$ 0.5 & 1.4\\
 MTS6 & 6 & 0.25 & 0.2 & 112.4 $\pm$ 1.8 & 37.9 $\pm$ 0.9 & 3.5\\
\end{tabular}
\end{ruledtabular}
\end{table*}
\bibliography{prr_soi}

\end{document}